\documentclass[10pt,letterpaper]{article}
\usepackage[table]{xcolor}
\usepackage[top=0.8in,left=1.5in,footskip=0.55in]{geometry}

\usepackage{amsmath,amssymb}

\usepackage{textcomp,marvosym}

\usepackage{cite}
\usepackage[draft]{changes}
\usepackage{nameref}
\usepackage[hidelinks]{hyperref}

\usepackage[left]{lineno}

\usepackage{microtype}
\DisableLigatures[f]{encoding = *, family = * }

\usepackage[table]{xcolor}

\usepackage{array}

\newcolumntype{+}{!{\vrule width 2pt}}

\newlength\savedwidth

\raggedright
\usepackage[aboveskip=1pt,labelfont=bf,labelsep=period,justification=raggedright,singlelinecheck=off]{caption}
\renewcommand{\figurename}{Fig}

\makeatletter
\renewcommand{\@biblabel}[1]{\quad#1.}
\makeatother

\usepackage{lastpage,fancyhdr,graphicx}
\usepackage{epstopdf}
\usepackage{graphicx}%
\usepackage{multirow}%
\usepackage{amsmath,amssymb,amsfonts}%
\usepackage{amsthm}%
\usepackage{mathrsfs}%
\usepackage{textcomp}%
\usepackage{manyfoot}%
\usepackage{makecell}%
\usepackage{cellspace}%
\usepackage{booktabs}%
\usepackage{listings}%
\usepackage{booktabs}%
\usepackage{makecell}%
\usepackage{color,soul}%
\usepackage{dsfont}%
\usepackage{tikz}%
\usepackage{textcomp}%

\usepackage{algorithm}
\usepackage{algpseudocode}

\usepackage{fixltx2e}

\MakeRobust{\Call}

\renewcommand{\algorithmiccomment}

\algnewcommand{\Inputs}[1]{%
  \State \textbf{Inputs:}
  \Statex \hspace*{\algorithmicindent}\parbox[t]{.8\linewidth}{\raggedright #1}
}
\algnewcommand{\Outputs}[1]{%
  \State \textbf{Outputs:}
  \Statex \hspace*{\algorithmicindent}\parbox[t]{.8\linewidth}{\raggedright #1}
}
\algnewcommand{\Initialize}[1]{%
  \State \textbf{Initialize:}
  \Statex \hspace*{\algorithmicindent}\parbox[t]{.8\linewidth}{\raggedright #1}
}
\renewcommand{\algorithmiccomment}[1]{\hfill$\triangleright$\protect{\footnotesize{\textit{#1}}}}

\def\ie{\textit{i.e.}}
\def\eg{\textit{e.g.}}

\usetikzlibrary{shapes.gates.logic.US}

\definecolor{gatecolor}{HTML}{0093D0}

\newcommand{\andgate}[1]{
  \raisebox{-0.3\height}{%
    \begin{tikzpicture}[transform shape, scale=#1]
      \node[and gate US, draw, logic gate inputs=nn, fill=gatecolor, fill opacity=0.6] {};
    \end{tikzpicture}%
  }
}
\newcommand{\orgate}[1]{
  \raisebox{-0.3\height}{%
    \begin{tikzpicture}[transform shape, scale=#1]
      \node[or gate US, draw, logic gate inputs=nn, fill=gatecolor, fill opacity=0.6] {};
    \end{tikzpicture}%
  }
}
\fancyheadoffset[L]{2.25in}
\fancyfootoffset[L]{1.25in}
\begin{document}
\vspace*{0.2in}

% Title must be 250 characters or less.
\begin{flushleft}
  {\Large \textbf\newline{Aggregating multiple test results to
      improve medical decision-making}
  % Please use "sentence case" for title and headings (capitalize only
  % the first word in a title (or heading), the first word in a
  % subtitle (or subheading), and any proper nouns).
}
\newline
% Insert author names, affiliations and corresponding author email (do not include titles, positions, or degrees).
\\
Lucas B\"ottcher\textsuperscript{1,*},
Maria R.\ D'Orsogna\textsuperscript{2,3},
Tom Chou\textsuperscript{3,4}
\\
\bigskip
\textbf{1} Dept.\ of Computational Science and Philosophy, Frankfurt School of Finance and Management, Frankfurt am Main, Germany
\\
\textbf{2} Dept.\ of Mathematics, California State University at Northridge, Los Angeles, CA, 91330, USA
\\
\textbf{3} Dept.\ of Computational Medicine, University of California, Los Angeles, Los Angeles, CA, 90095-1766, USA
\\
\textbf{4} Dept.\ of Mathematics, University of California, Los Angeles, Los Angeles, CA, 90095-1555, USA
\\
\bigskip
% Insert additional author notes using the symbols described below. Insert symbol callouts after author names as necessary.
% 
% Remove or comment out the author notes below if they aren't used.
%
% Primary Equal Contribution Note
% \Yinyang These authors contributed equally to this work.

% Additional Equal Contribution Note
% Also use this double-dagger symbol for special authorship notes, such as senior authorship.
% \ddag These authors also contributed equally to this work.

% Current address notes
% \textcurrency Current Address: Dept/Program/Center, Institution Name, City, State, Country % change symbol to "\textcurrency a" if more than one current address note
% \textcurrency b Insert second current address 
% \textcurrency c Insert third current address

% Deceased author note
% \dag Deceased

% Group/Consortium Author Note
% \textpilcrow Membership list can be found in the Acknowledgments section.

% Use the asterisk to denote corresponding authorship and provide email address in note below.
* l.boettcher@fs.de

\end{flushleft}

%\linenumbers

% Please keep the abstract below 300 words
\section*{Abstract}
Gathering observational data for medical decision-making
often involves uncertainties arising from both type I (false positive)
and type II (false negative) errors. In this work, we develop a
statistical model to study how medical decision-making can be improved
by repeating diagnostic and screening tests, and aggregating their results. This
approach is relevant not only in clinical settings, such as medical
imaging, but also in public health, as highlighted by the need
for rapid, cost-effective testing methods during the SARS-CoV-2
pandemic. Our model enables the development of testing protocols with an arbitrary number of tests, which can be customized to meet requirements for type I and type II errors. This allows us to adjust sensitivity and specificity according to application-specific needs. Additionally, we derive generalized Rogan--Gladen estimates for estimating disease prevalence, accounting for an arbitrary number of tests with potentially different type I and type II errors. We also provide the corresponding uncertainty quantification.
% Please keep the Author Summary between 150 and 200 words
% Use first person. PLOS ONE authors please skip this step. 
% Author Summary not valid for PLOS ONE submissions.   
\section*{Author summary}
Our work focuses on medical decision-making, particularly on
addressing uncertainties associated with screening and diagnostic
tests. No test is perfect, so finding a balance between false
positives (misidentifying a condition) and false negatives (missing a
condition) is crucial in many biomedical applications.  Implementing
accurate and efficient testing is important not only for individual
diagnoses but also for population-wide testing during a pandemic.
Since cost-effective and rapid tests are often quite inaccurate, a
common goal is to obtain accurate assessments from repeated testing
and meaningfully combining their results. However, using the
multitude of tests and their different sequences of administration to
design effective test protocols is a challenge that requires new
statistical tools.  In this study, we develop tools for
  aggregating test results in ways that can be tailored to specific
applications by tuning the false positive-false negative
ratio. Furthermore, we demonstrate how our method can improve disease
prevalence estimates and thus aid in the implementation of effective
public health measures.

% Use "Eq" instead of "Equation" for equation citations.
\section*{Introduction}
Administering effective diagnostic and screening tests plays an important role in
most biomedical decision-making. Recent advancements in biotechnology have made
a wide array of biochemical tests readily available on a
large scale. For example, in the case of SARS-CoV-2, a systematic review identified
49 different antigen tests~\cite{dinnes2022rapid} which are
cost-effective and can provide results in 15--30 minutes.
However, their sensitivity (\ie, true positive rate) can be as low as 34.3\% in
symptomatic patients and 28.6\% in asymptomatic
patients~\cite{dinnes2022rapid}. This indicates that some tests
correctly identify an infected individual as positive in only about one third of
cases, leaving a significant portion of those with the 
disease undetected. Besides sensitivity, another metric used to assess
the accuracy of a test is its specificity (\ie, true
negative rate).  Highly sensitive tests prioritize identifying individuals with a disease, while highly specific tests prioritize identifying those who do not have the disease. In most cases, sensitivity and specificity are inversely related; 
both are important when assessing the value of a medical test \cite{glaros1988understanding, akobeng2007understanding}. 

Given the availability of various tests with differing sensitivities
and specificities, how can one repeat tests and integrate results to
minimize both type I errors (false positives) and type II errors
(false negatives)? Although this is a key question across many
different clinical settings, including diabetes
testing~\cite{brohall2006prevalence,kermani2017accuracy}, medical
imaging~\cite{weinstein2005clinical,zou2006statistical,brennan2019benefits}, prostate cancer testing~\cite{vickers2013}, and stool sample analysis in colon cancer
testing~\cite{neuhauser1975we,collins2005accuracy}, our primary focus
will be on aggregating results from different tests within the
context of SARS-CoV-2 due to the availability of comprehensive studies
on properties of the corresponding tests.

The SARS-CoV-2 pandemic emphasized the crucial role of testing in managing the spread 
of an infectious disease.
During the early stages of the pandemic, test shortages were common, causing delays in diagnoses, underreporting of COVID-19 cases, and hindering the effectiveness of public health measures.
Due to the intensifying crisis, regulatory agencies expedited the review and approval process 
of dedicated tests developed by different suppliers in different countries that used different technologies. 
These tests were often in use
at the same time.\footnote{One
  distinguishes between two primary categories of SARS-CoV-2 tests:
  (i) viral tests and (ii) antibody (or serological)
  tests~\cite{cdc_sarscov2testing}. Within the viral test category,
  there exist two main subclasses: nucleic acid amplification tests
  (NAATs), such as reverse transcription polymerase chain reaction
  (RT-PCR) tests that typically detect viral RNA, 
  and antigen tests that detect specific antigen proteins on the surface of the virus.
  Antibody tests serve to identify antibodies produced as
  part of the adaptive immune system response. In the context of
  SARS-CoV-2, antibody tests may target anti-nucleocapsid antibodies,
  indicative of current or past infection, and anti-spike protein
  antibodies, generated through infection or vaccination.}
Early detection methods relied on 
genetic sequencing and reverse transcription polymerase chain reaction (RT-PCR) tests to detect
viral genetic material. 
Antibody tests were also introduced to detect the presence of the virus in
previously infected individuals who had developed an immune response.
Both tests used samples collected from nasopharyngeal swabs and required
specialized laboratory equipment and personnel to process them, 
making the diagnosis of active infections (RT-PCR tests) or of an activated immune response
(antibody tests) available only after a few hours or even days. 
As the pandemic surged, the prioritization of rapid testing methods led to the development of rapid antigen tests, capable of detecting viral proteins and providing results within minutes.
Subsequent saliva-based tests offered a less invasive experience compared
to those based on nasopharyngeal swabs. Finally, the retreat of the pandemic was accompanied by the
introduction of home testing kits. Current research is focused on perfecting new methods, including breathalyzer tests and wastewater monitoring. 

 Each testing method
has its specific advantages and limitations. For example, RT-PCR tests are highly sensitive and specific and can detect even small amounts of viral RNA. However, there may be long delays in obtaining actionable results. 
Antibody tests may not detect antibodies in the early stages of the infection
and are prone to large false-positive results due to cross-reactivity with antibodies from other 
viruses. Antigen tests are usually less sensitive than RT-PCR tests, but yield the
most rapid response. Further variability in sensitivity and specificity arises within each type of testing method due to differences among test manufacturers and periodic modifications to the biochemical protocols, which are made to ensure the detection of any novel viral mutations or variants.

Our collective past experience with the spread of the SARS-CoV-2 virus poses several 
preparedness challenges to better respond to future pandemics, including how to best allocate scarce resources and 
enhance testing and classification strategies. 
The development of appropriate mathematical and computational methods plays a fundamental
role in addressing these challenges. For example, one way to stretch resources is to test pooled samples,
allowing one to eliminate large numbers of uninfected individuals with a small number of tests. 
Several mathematical approaches have been developed to study the optimization of both sample pooling and testing \cite{pool2021}. Different proposed strategies depend on test sensitivity and specificity
\cite{safe_pooling}, estimated prevalence \cite{pooling_low_prevalence,pooling_high_prevalence}, disease
dynamics \cite{pooling_VL_optimized}, and available social contact information \cite{adaptive_pooling}. 
Other mathematical approaches aimed at improving testing efficiency 
account for  uncertainty in disease
prevalence \cite{patrone2021classification}, indeterminate test
results \cite{patrone2022optimal}, time-dependent prevalence and
antibody levels \cite{bedekar2023prevalence,bedekar2024},
high-dimensional data analysis to improve classification accuracy
\cite{luke2023modeling}, and multiple classes such as vaccinated,
previously infected, and unexposed individuals \cite{luke2023optimal}.

In this paper, we focus on developing mathematical and computational methods that can help improve medical decision-making by repeating tests and aggregating their results. Several related studies have highlighted the potential of this approach~\cite{cebul1982using,mcclish1985improving,hershey1986clinical,marshall1989predictive,ramdas2020test,jain2021new,jain2023robust,salvatore2022quantitative,perkmann2023increasing},
often using different terms such as ``all
heuristic''~\cite{jain2021new,jain2023robust}, ``believe-the-negative
rule''~\cite{politser1982reliability}, ``conjunctive positivity criterion''~\cite{hershey1986clinical,ament1993,felder2022}, and ``orthogonal
testing''~\cite{lau2021disease} to refer to the same protocol
where all tests must return a positive result in order to classify an
individual as infected. In Boolean algebra, this corresponds to an
aggregation using the binary $\mathrm{AND}$ operator. 
Another aggregation method is the ``any heuristic''~\cite{jain2021new,jain2023robust} 
also termed the ``believe-the-positive rule''~\cite{politser1982reliability} or ``disjunctive positivity criterion''~\cite{hershey1986clinical,ament1993,felder2022}. 
In this protocol, all tests must return a negative result in
order to classify an individual as not infected. It is thus  sufficient for one
test to be positive for a positive diagnosis. In Boolean algebra, this 
aggregation method is represented by the binary $\mathrm{OR}$ operator.  

The US Food and Drug Administration (FDA) has also recognized the
relevance of repeated testing and released an Excel-based calculator
to compute properties of two combined tests~\cite{FDAcalculator}. 
However, most available result aggregation methods, including the FDA calculator,
only consider two tests and usually employ very few (between one and
three) basic aggregation methods. Nevertheless, there are instances
where jurisdictions have implemented testing protocols involving three
and four tests, such as in Vienna, Austria~\cite{breyer2021low}, and Santiago,
Chile~\cite{VIAL2022100606}. 
Without appropriate mathematical insight and computational tools, however,
it is challenging to analyze the properties of all possible
aggregation methods due to the vast number of tests and their
combinations. The lack of theoretical understanding 
often results in the implementation of ad-hoc and
suboptimal aggregation protocols, rather than the most efficient ones. 
In addition to determining the disease status of an individual,
combined tests can improve estimates of disease
prevalence~\cite{rogan1978estimating,mcclish1985improving}, which is
helpful in infectious-disease surveillance and
management~\cite{bottcher2021using,zhang2022data,schneider2022epidemic,bottcher2022statistical,meyer2022adjusting,felder2022,zhou2023correcting,owusu2023dynamics,levin2022assessing,Liu2023}. In this context, it is also important to develop suitable mathematical tools to compare disease prevalence estimates across jurisdictions, as different public health organizations employ different testing protocols and aggregation methods~\cite{poljak2021seroprevalence,seroprevalence-norrbotten,breyer2021low,10.1093/cid/ciac198,VIAL2022100606}.

Here, we combine concepts from biostatistics and Boolean
algebra to develop a broadly applicable statistical model that
  can guide medical decision-making after repeated screening or
diagnostic testing. We show how our model enables the development
of testing protocols whose overall sensitivity and
  specificity can be tuned to satisfy application-specific
  requirements on type I and type II errors. Additionally, we present an algorithm capable of determining the best way to aggregate results from a given set of tests in terms of efficient sensitivity-specificity pairs.
Furthermore, we integrate our
  aggregation approach with population-level prevalence estimation, demonstrating how repeated testing can enhance prevalence monitoring. Specifically, we generalize the Rogan--Gladen prevalence estimate  ~\cite{rogan1978estimating,mcclish1985improving} to account for an arbitrary number of tests, each with potentially
  different type I and type II errors.
\section*{Results}
\subsection*{Aggregating two tests}
\label{sec:two_tests}
\begin{figure}[t!]
    \centering
    \includegraphics[width=0.9\textwidth]{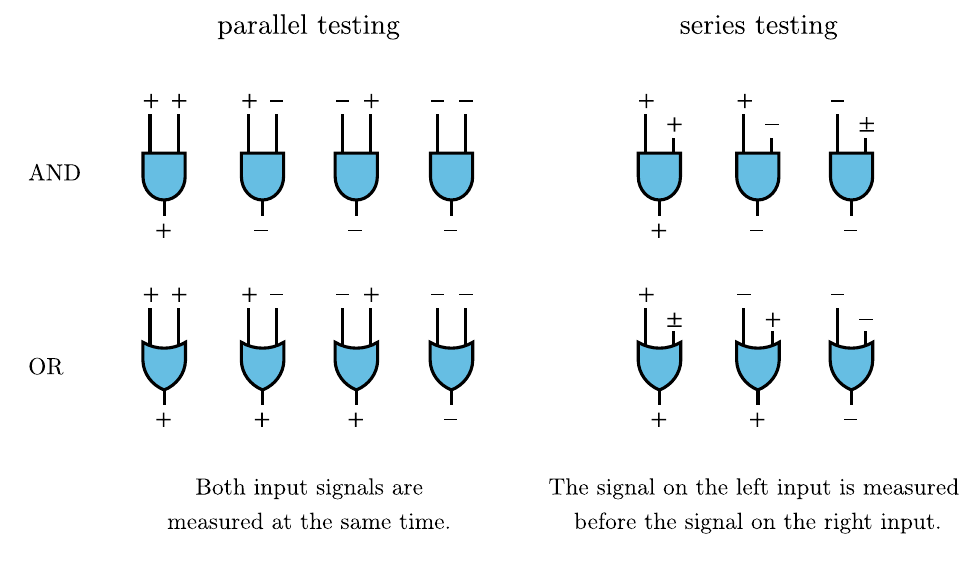}
    \caption{Parallel and series testing protocols using two
      tests. Positive ($+$) and negative ($-$) test outcomes are
      combined using the two Boolean functions $\mathrm{AND}$
      (\protect\andgate{0.5}) and $\mathrm{OR}$
      (\protect\orgate{0.5}). In parallel testing, both inputs are
      assessed simultaneously, while in series testing, the left input
      is examined before the right. Hence, if the initial test in a
      series protocol yields a negative result with aggregation
      through an $\mathrm{AND}$ gate, the final result will be
      negative, irrespective of the second input. In series testing
      with an $\mathrm{OR}$ gate, the overall result will be positive
      if the first test is positive, regardless of the outcome of the
      second test.}
    \label{fig:parallel_serial_two}
\end{figure}
As a starting point, we examine a testing protocol that combines the
results of $n=2$ tests (possibly of different types), denoted by
binary random variables $Y_1$ and $Y_2$, where $Y_1, Y_2 \in
\{0,1\}$. The disease status of an individual, classified as
either positive ($+$) or negative ($-$), is represented by another
binary random variable $X \in \{0,1\}$.

The true positive rates TPRs (or sensitivities) of each of the
  two (type 1 and type 2) tests are defined as
\begin{align}
    {\rm TPR}_{1}=\Pr(Y_1=1\mid X=1)\quad\text{and}\quad  {\rm TPR}_{2}=\Pr(Y_2=1\mid X=1)\,,
\end{align}
respectively. The corresponding true negative rates TNRs (or specificities) are
\begin{align}
    {\rm TNR}_{1}=\Pr(Y_1=0\mid X=0)\quad\text{and}\quad {\rm TNR}_{2}=\Pr(Y_2=0\mid X=0)\,,
\end{align}
respectively. These individual-test TPRs and TNRs serve as
  building blocks for modeling the overall TPR and TNR of a testing
  and aggregation protocol involving multiple tests.

For $n=2$ ordered tests, there are $r=2^n=2^2=4$ possible
sequences of test results (\ie, the inputs are permutations of ``$+$''
and ``$-$'' of length 2): $(+,+)$, $(+,-)$, $(-,+)$, and $(-,-)$. Given that the final, aggregated output can be either ``$+$''
  or ``$-$'', the number of possible output sequences that one can
  assign to each of the $r=4$ input configurations is $2^{r} = 16$,
  which is equivalent to the total number of two-input Boolean gates.

In Figure~\ref{fig:parallel_serial_two} shows two (out of 16)
possible output sequences associated with the aggregation of test
results that involve $n=2$ tests. In this example, aggregation can be
processed through an $\mathrm{AND}$ gate or an $\mathrm{OR}$ gate. Depending on how the individual test results are processed, the output of the chosen aggregation function yields a positive or negative overall result. Notice that the aggregated output sequences of the parallel and series testing protocols shown are the same. The only implicit difference lies in how the input signals are processed (parallel or
  series). Both $\mathrm{AND}$ and
$\mathrm{OR}$ aggregation functions have been used in SARS-CoV-2
seroprevalence studies (see Table~\ref{tab:examples}) and we
  will analyze both in this paper.

Aggregating test results using an $\mathrm{AND}$ gate produces a
positive result if and only if both inputs are positive, corresponding to a ``conjunctive positivity criterion''~\cite{hershey1986clinical,ament1993,felder2022}. Otherwise the output result is negative. For an $\mathrm{OR}$ gate,
the aggregate test is negative if and only if both inputs are
negative, corresponding to a ``disjunctive positivity criterion''~\cite{hershey1986clinical,ament1993,felder2022}. In the remaining cases, the $\mathrm{OR}$ aggregation method
outputs a positive result. The process of aggregation is
  sometimes referred to as the ``all  heuristic''~\cite{jain2021new,jain2023robust} and the
  ``believe-the-negative rule''~\cite{politser1982reliability} when
  using $\mathrm{AND}$ aggregation. Similarly, $\mathrm{OR}$
  aggregation is sometimes termed the ``any
  heuristic'~\cite{jain2021new,jain2023robust} and the
  ``believe-the-positive rule''~\cite{politser1982reliability}. For the two possible test-administration orderings
  (parallel and series) and the two aggregation procedures
  ($\mathrm{AND}$ and $\mathrm{OR}$ gates), we denote the corresponding cases as
  series $\mathrm{AND}$, series $\mathrm{OR}$, parallel
  $\mathrm{AND}$, and parallel $\mathrm{OR}$.

We use the random variable $Z\in\{0,1\}$ to denote the aggregated
output and first examine the parallel testing protocol with an
$\mathrm{AND}$ aggregation function. For $n=2$ parallel tests, the
sensitivity and specificity are
\begin{equation}
\begin{aligned}
	{\rm TPR}^{(\mathrm{p})}_{1\land 2} = & \Pr(Z=1\mid X=1) 
	   \label{eq:tpr12_par_and} =  \Pr(Y_1=1,Y_2=1\mid X=1) =  {\rm TPR}_{1} {\rm TPR}_{2}
\end{aligned}
\end{equation}
and

\begin{equation}
\begin{aligned}
	       \label{eq:tnr12_par_and}
  {\rm TNR}^{(\mathrm{p})}_{1\land 2} & =\Pr(Z=0\mid X=0) \\
     & =  1 - \Pr(Y_1=1,Y_2=1\mid X=0) \\
   & =  \Pr(Y_1=0,Y_2=0\mid X=0)+\Pr(Y_1=0,Y_2=1\mid X=0)+\Pr(Y_1=1,Y_2=0\mid X=0)\\
    & = {\rm TNR}_{1} {\rm TNR}_{2}+{\rm TNR}_{1}(1-{\rm TNR}_{2})+{\rm TNR}_{2}(1-{\rm TNR}_{1})\\
    & = {\rm TNR}_{1}+(1-{\rm TNR}_{1}){\rm TNR}_{2}\,,
\end{aligned}
\end{equation}
respectively. In Eqs.~\eqref{eq:tpr12_par_and} and
\eqref{eq:tnr12_par_and}, we assumed that the results of different
tests are conditionally independent given the disease status.

Conditioned on a disease state, this independence assumption is
  reasonable provided the tests do not interfere with each other,
  chemically or through biases in administration and interpretation,
  or perturb the patient. For example, the assays used in 
  Elecsys$^\text{\textregistered}$ Anti-SARS-CoV-2 (Elecsys-N) and
  Elecsys$^\text{\textregistered}$ Anti-SARS-CoV-2 S (Elecsys-S)
  target antibodies against different proteins of the virus,  nucleocapsid and spike proteins, respectively, 
so that  test interactions can be
  neglected. Repeated tests that involve interpreting images can instead carry
  correlations due to changing operator bias. This phenomenon has been studied in
  \cite{zou2006statistical}.

The derivations presented above are applicable to $n=2$ parallel
tests, where both the first \emph{and} the second test results must be
positive for classifying a sample as positive (an $\mathrm{AND}$
gate). If, however, the classification is based on the first \emph{or}
the second result being positive (an $\mathrm{OR}$ gate), the
sensitivity and specificity are
\begin{equation}
  \begin{aligned}
    {\rm TPR}^{(\mathrm{p})}_{1\lor 2} & =  \Pr(Z=1\mid X=1) \\
         & = 1 - \Pr(Y_{1}=0, Y_{2}=0\mid X=1) \\
            & = \Pr(Y_{1}=1, Y_{2}=1\mid X=1) + \Pr(Y_{1}=1, Y_{2}=0\mid X=1)
    + \Pr(Y_{1}=0, Y_{2}=1\mid X=1) \\
    & =  {\rm TPR}_{1} {\rm TPR}_{2} +  {\rm TPR}_{1}(1- {\rm TPR}_{2})
    + {\rm TPR}_{2}(1-{\rm TPR}_{1}) \\
     & = {\rm TPR}_{1}+(1-{\rm TPR}_{1}){\rm TPR}_{2} \\
\end{aligned}
\label{eq:tpr12_par_or}
\end{equation}
and
\begin{equation}
    {\rm TNR}^{(\mathrm{p})}_{1\lor 2}=  \Pr(Z=0\mid X=0) 
    = \Pr(Y_{1}=0, Y_{2}=0\mid X=0)=  {\rm TNR}_{1} {\rm TNR}_{2}\,,
\label{eq:tnr12_par_or}
\end{equation}
respectively. Given the assumptions in deriving the AND and OR aggregation protocols, we expect
that the true positive rate is lower under AND aggregation (since all tests must be positive
for a positive diagnosis) and vice-versa that the true negative rate is lower under OR
aggregation (since all tests must be negative for a negative diagnosis). 
Based on Eqs.~\eqref{eq:tpr12_par_and}--\eqref{eq:tnr12_par_or}, we obtain
\begin{equation}
  {\rm TPR}^{(\mathrm{p})}_{1\lor 2} \geq  {\rm TPR}^{(\mathrm{p})}_{1\land 2}
  \quad \mbox{and}\quad  {\rm TNR}^{(\mathrm{p})}_{1\lor 2}
  \leq {\rm TNR}^{(\mathrm{p})}_{1\land 2}
  \end{equation}
for all $\mathrm{TPR}_i$ and $\mathrm{TNR}_i$ ($i\in\{1,2\}$).

Instead of administering two tests in parallel, one may also consider
series testing in which the second test is administered depending on
the outcome of the first test. In contrast to parallel testing with an
$\mathrm{AND}$ aggregation, the second test in the corresponding
sequential testing protocol does not have to be performed if the outcome
of the first test is negative. The sensitivity and specificity of
series testing under $\mathrm{AND}$ aggregation are
\begin{equation}
  {\rm TPR}^{(\mathrm{s})}_{1\land 2}={\rm TPR}_{1} {\rm TPR}_{2}
  \quad\text{and}\quad {\rm TNR}^{(\mathrm{s})}_{1\land 2}={\rm TNR}_{1}
  + (1-{\rm TNR}_{1}){\rm TNR}_{2}\,,
    \label{eq:tpr12_tnr12_ser_and}
\end{equation}
respectively. For the corresponding series $\mathrm{OR}$ test, we have
\begin{equation}
  {\rm TPR}^{(\mathrm{s})}_{1\lor 2}={\rm TPR}_{1} + (1-{\rm TPR}_{1}){\rm TPR}_{2}
  \quad\text{and}\quad {\rm TNR}^{(\mathrm{s})}_{1\lor 2}={\rm TNR}_{1}{\rm TNR}_{2}\,.
    \label{eq:tpr12_tnr12_ser_or}
\end{equation}
Notice that the sensitivities and specificities of the aggregated
tests are the same regardless of whether a parallel or sequential
aggregation protocol is employed.  However, in a sequential protocol,
fewer tests need to be administered, making this option more
economically viable, especially for rapid antigen tests, characterized by 
lower sensitivity.  For tests
with extended processing times, such as enzyme-linked immunosorbent
assay (ELISA) and RT-PCR tests, one may still prefer parallel test
protocols to avoid substantial delays between the first and second
tests.

Mathematically, the TPRs and TNRs of the studied combined testing protocols bound the TPRs and TNRs of the constituent tests according to
\begin{equation}
  {\rm TPR}_{1\land 2} \leq {\rm TPR}_{i} \leq  {\rm TPR}_{1\lor 2}\quad \mbox{and}
  \quad {\rm TNR}_{1\lor 2} \leq {\rm TNR}_{i} \leq {\rm TNR}_{1\land 2}\quad {\mbox {for}}\quad i\in\{1,2\}\,.
    \label{eq:inequalities}
\end{equation}
We will show that this bounding result also holds for $n\geq 3$ tests.
\subsubsection*{Saving tests with series testing}
\begin{figure}[t!]
    \centering
    \includegraphics[width=\textwidth]{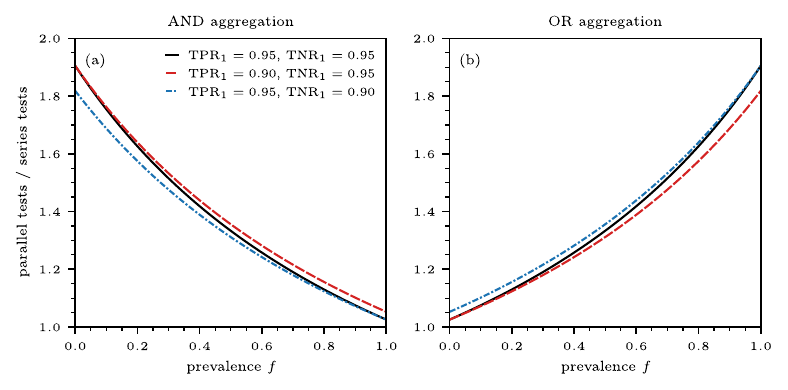}
    \caption{The ratio of the number of parallel tests
      to the number of series tests necessary to determine the aggregated
      output from $n=2$ tests as a function
      of prevalence $f$. 
       Results in panels (a) and
      (b) are based on $\mathrm{AND}$ and $\mathrm{OR}$ aggregations
      of two tests, using Eqs.~\eqref{ratio1} and \eqref{ratio2}, respectively. We consider three different
      combinations of true positive and true negative rates (solid
      black lines: $\mathrm{TPR}_1=0.95$ and $\mathrm{TNR}_1=0.95$;
      dashed red lines: $\mathrm{TPR}_1=0.90$ and
      $\mathrm{TNR}_1=0.95$; dash-dotted blue lines:
      $\mathrm{TPR}_1=0.95$ and $\mathrm{TNR}_1=0.90$).
      The critical values $f_{\rm c}$ for which the ratios in panel (a) are larger than the ratios in
      panel (b) are given, respectively, by $f_{\rm c} =0.50,  0.47, 0.53$. For $f < f_{\rm c}$
      greater savings are achieved by utilizing the $\rm AND$-aggregated series tests, compared to the $\rm OR$-aggregated
      series test.}
    \label{fig:parallel_vs_serial}
\end{figure}
To fully cover a population comprising $N$ individuals using parallel
testing would require $2N$ tests. In contrast, series testing involves administering an initial test to all individuals. In the series $\mathrm{AND}$ aggregation function, a second test is required
if and only if the first test yields a positive result. The probability of this
event is $f {\rm TPR}_1 + (1-f) (1-{\rm TNR}_1)$, where $f\in[0,1]$ is the prevalence, the fraction of the total population carrying
a disease. In the series $\mathrm{OR}$ aggregation function, a
second test is necessary if and only if the first test is negative,
and the probability of this event is $f (1-{\rm TPR}_1) + (1-f) {\rm
  TNR}_1$. Both series testing protocols achieve the same sensitivity and
specificity as the parallel test but with fewer tests, 
specifically, $N(1+f {\rm TPR}_1 + (1-f) (1-{\rm TNR}_1))$ tests for
the series $\mathrm{AND}$ function and $N(1 + f (1-{\rm TPR}_1) +
(1-f) {\rm TNR}_1)$ tests for the series $\mathrm{OR}$ function,
instead of $2N$ when conducted in parallel. 

We assume the number of tests available is such that
there are enough of them to cover the entire population $N$.
Since under parallel testing all individuals must be tested at least twice, 
this assumption implies that the population is at the most half the number of available tests. 
Given the disease prevalence $f$, we now compute the ratio of the number of required tests
under parallel testing and the corresponding number of sequential tests.
The ratios for the $\mathrm{AND}$ and $\mathrm{OR}$ aggregation methods are
\begin{equation}
 \frac{\rm {parallel \,tests}}{\rm {series \, tests}} 
\bigg  \rvert_{1 \land 2} =  \frac{2} {1 + f {\rm TPR}_1 + (1-f) (1-{\rm TNR}_1)}
\label{ratio1}
\end{equation}
and
\begin{equation}
 \frac{\rm {parallel \,tests}}{\rm {series \, tests}} 
\bigg  \rvert_{1 \lor 2} =  \frac{2}{1 + f (1-{\rm TPR}_1) + (1-f) {\rm TNR}_1}\,,
\label{ratio2}
\end{equation}
respectively. Both ratios lie between 1 and 2, 
indicating that parallel testing always requires more tests
than series testing. Besides the ground truth prevalence $f$,
  this ratio also depends on the disposition of the first test that
  determines if a second test is warranted.  The first test result in
  turn, depends on its sensitivity TPR$_1$ and specificity TNR$_1$.
  Figure~\ref{fig:parallel_vs_serial} shows these ratios as a function of
  prevalence $f$ for three different combinations of true positive and
true negative rates: (i) $\mathrm{TPR}_1=0.95$ and
$\mathrm{TNR}_1=0.95$, (ii) $\mathrm{TPR}_1=0.90$ and
$\mathrm{TNR}_1=0.95$, and (ii) $\mathrm{TPR}_1=0.95$ and
$\mathrm{TNR}_1=0.90$.  When there are no infected individuals in the
population (\ie, $f=0$), the parallel to series ratios are
$2/(2-\mathrm{TNR}_1)$ and $2/(1+\mathrm{TNR}_1)$ for the
$\mathrm{AND}$ and $\mathrm{OR}$ aggregation schemes, respectively. If
all $N$ individuals in a population are infected (\ie, $f=1$), the
ratios are $2/(1+\mathrm{TPR}_1)$ and $2/(2-\mathrm{TPR}_1)$ for the
$\mathrm{AND}$ and $\mathrm{OR}$ aggregation schemes, respectively.

It is also straightforward to show that for $f < f_{\rm c}$,  where $f_{\rm c}$
is the critical prevalence defined as

\begin{equation}
\label{fcrit}
  f_{\rm c} = \frac{2\mathrm{TNR}_1-1}{2(\mathrm{TPR}_1+\mathrm{TNR}_1-1)}, 
\end{equation}
the number of required $\mathrm{AND}$-aggregated series tests is less than 
the number of required $\mathrm{OR}$-aggregated series tests. Equivalently, 
for $f < f_{\rm c}$, the curve representing
the ratio of the required parallel to series tests under the $\mathrm{AND}$ protocol
given in Eq.~\eqref{ratio1} 
falls above the corresponding $\mathrm{OR}$ protocol
curve given in Eq.~\eqref{ratio2}. The trends observed in the parallel-to-series ratios as a function of prevalence $f$, shown in Figure~\ref{fig:parallel_vs_serial}(a), confirm that $\rm AND$ aggregation yields greater test savings through series testing for prevalences $f<f_{\rm c}$. In contrast, $\rm OR$ aggregation results in larger savings for $f>f_{\rm c}$, as illustrated in Figure~\ref{fig:parallel_vs_serial}(b). 

According to Eq.~\eqref{fcrit}, the quantity $f_{\rm c}$ is meaningful only when 
TNR$_{1} \geq 1/2$ and when $\mathrm{TPR}_1+\mathrm{TNR}_1 > 1$. The latter
condition implies that the true
  positive rate of the first test is greater than its false positive
  rate (\ie, $1-\mathrm{TNR}_1$). A test that satisfies this condition is said to have ``discriminatory power''~\cite{felder2022}.
  Typical values of $\mathrm{TPR}_1$ and $\mathrm{TNR}_1$ yield
  intermediate values of $f_{\rm c} \approx 0.5$ as shown  in Figure~\ref{fig:parallel_vs_serial}. 
  Another scenario in which $f_{\rm c}$ is mathematically meaningful is when 
  TNR$_{1} \leq 1/2$ and $\mathrm{TPR}_1+\mathrm{TNR}_1 < 1$. 
  In this case, the trends in Figures~\ref{fig:parallel_vs_serial}(a,b) are
  reversed compared to the ones just discussed. This scenario, however, is highly unrealistic, 
  as the first test is misleading since its false positive
  rate is greater that its true positive rate.
\subsubsection*{Positive predictive value}
\begin{figure}[t!]
    \centering
    \includegraphics[width=\textwidth]{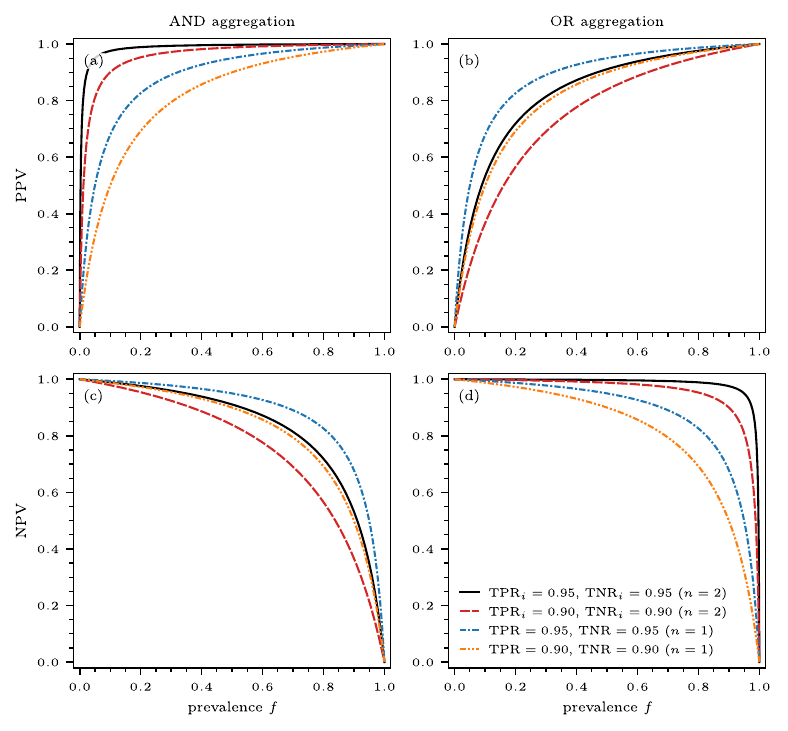}
    \caption{Positive predictive value (PPV) and negative predictive
      value (NPV) as a function of prevalence $f$. The results that we
      show in panels (a,c) and (b,d) are based on $\mathrm{AND}$ and
      $\mathrm{OR}$ aggregations of $n=2$ tests, using Eqs.\,\eqref{PPVdef} and \eqref{NPVdef}, respectively. We
      denote the sensitivities and specificities of the two tests $i\in\{1,2\}$ by
      $\mathrm{TPR}_i$ and $\mathrm{TNR}_i$,
      respectively. We consider two different combinations of true
      positive and true negative rates (solid black lines:
      $\mathrm{TPR}_i=0.95$ and $\mathrm{TNR}_i=0.95$; dashed red
      lines: $\mathrm{TPR}_i=0.90$ and $\mathrm{TNR}_i=0.90$). As a
      reference, we also show results for single tests without further
      aggregation (dash-dotted blue line: $\mathrm{TPR}=0.95$ and
      $\mathrm{TNR}=0.95$; dash-dot-dotted orange line:
      $\mathrm{TPR}=0.90$ and $\mathrm{TNR}=0.90$). These
        curves are independent of the ordering (parallel or series)
        method used.}
    \label{fig:ppv_vs_npv}
\end{figure}
Measures such as sensitivity and specificity fail to
appropriately take into account the prevalence of a disease $f$~\cite{eisenberg1995accuracy}. In this context, a more
appropriate measure is the positive predictive value (PPV), also known
as precision, defined as
\begin{equation}
\label{PPVdef}
    \mathrm{PPV} = \frac{f \mathrm{TPR}}{f \mathrm{TPR} + (1-f)(1-\mathrm{TNR})}\,.
\end{equation}
The $\mathrm{PPV}$ is the number of true positives divided by the
number of positive calls.  Similarly, we define the negative
predictive (NPV) value as the number of true negatives divided by the
number of negative calls,
\begin{equation}
\label{NPVdef}
  \mathrm{NPV} = \frac{(1-f) \mathrm{TNR}}{ (1-f) \mathrm{TNR}
    + f (1-\mathrm{TPR})}\,.
\end{equation}
Here, $\mathrm{TPR}$ and $\mathrm{TNR}$ represent the overall
  true positive and true negative rates of the aggregate testing
  protocol. 
  By defining the utility gain associated with treating a sick individual and the utility loss associated with treating a healthy individual, it is possible to establish a relationship between $\mathrm{PPV}$, $\mathrm{NPV}$, and the treatment threshold. This threshold is the point where the expected treatment gain equals the expected treatment loss~\cite{felder2022}.
 
  Based on Eqs.~\eqref{PPVdef} and \eqref{NPVdef}, one can show that the PPV is an increasing function of $f$
 and that the NPV is a decreasing function of $f$. These equations also yield $\mathrm{PPV} \geq \mathrm{NPV}$
  when
\begin{equation}
  f \geq \frac{\sqrt{\mathrm{TNR}\big(1-\mathrm{TNR}\big)}}{
    \sqrt{\mathrm{TNR}\big(1-\mathrm{TNR}\big)} +
    \sqrt{\mathrm{TPR}\big(1-\mathrm{TPR}\big)}}\,.
\end{equation}
For multiple tests, the PPV and NPV are independent of the test
  ordering (parallel or series); however, they depend on the TPRs and
  TNRs of the individual tests in different ways depending whether the
  $\mathrm{AND}$ or the $\mathrm{OR}$ aggregation protocol is
  used. Specifically, we have 
\begin{equation}
  \mathrm{PPV}_{1\land 2} \geq \mathrm{PPV}_{1\lor 2}\,, \qquad \forall f 
  \label{inequalities}
  \end{equation}
 if
\begin{equation}
\label{ppvcond}
\frac{\mathrm{TPR}_1 +\mathrm{TPR}_2 }{ \mathrm{TPR}_1 \mathrm{TPR}_2} \leq 
\frac{\mathrm{(1-TNR}_1) +(1 - \mathrm{TNR}_2) }{ ( 1- \mathrm{TNR}_1)(1- \mathrm{TNR}_2)}\,.
\end{equation}
Similarly, we find
  
  \begin{equation}
  \mathrm{NPV}_{1\land 2} \leq \mathrm{NPV}_{1\lor 2}\,, \qquad \forall f 
  \label{inequalities2}
  \end{equation}
if
\begin{equation}
\label{npvcond}
\frac{\mathrm{TNR}_1 +\mathrm{TNR}_2 }{ \mathrm{TNR}_1 \mathrm{TNR}_2} \leq 
\frac{\mathrm{(1-TPR}_1) +(1 - \mathrm{TPR}_2) }{ ( 1- \mathrm{TPR}_1)(1- \mathrm{TPR}_2)}. \
\end{equation}
The conditions in Eqs.~\eqref{ppvcond} and \eqref{npvcond} are always satisfied if tests with discriminatory power are used, {\textit {i.e.}} if $\mathrm{TPR}_{1}+\mathrm{TNR}_{1} \geq 1$ and
  $\mathrm{TPR}_{2}+ \mathrm{TNR}_{2} \geq 1$. 
  
  In Figure~\ref{fig:ppv_vs_npv}, we show the dependence of
$\mathrm{PPV}$ and $\mathrm{NPV}$ on the prevalence $f$, using
sensitivities and specificities associated with $\mathrm{AND}$ and
$\mathrm{OR}$ aggregations for two tests [see   Eqs.~\eqref{eq:tpr12_par_and}--\eqref{eq:tnr12_par_or}]. We also include the corresponding
$\mathrm{PPV}$ and $\mathrm{NPV}$ of individual (unaggregated) tests
for reference.  As can be seen,  tests aggregated with the
$\mathrm{AND}$ function yield substantially higher PPVs compared to those aggregated
with an $\mathrm{OR}$ function for all $f$, while  the $\mathrm{OR}$ aggregation
results in notably higher NPVs compared to those obtained with the $\mathrm{AND}$ aggregation. Figure~\ref{fig:ppv_vs_npv} also shows that for low prevalence $f$
the highest PPV and NPV values are obtained under $\mathrm{AND}$ aggregation, 
whereas $\mathrm{OR}$ aggregation is best for high prevalence $f$. 
For low $f$, the main source of test error is the false positive rate $1 - \mathrm{TNR}$. This term is minimized under the $\mathrm{AND}$ aggregation as per 
Eq.~\eqref{eq:inequalities}. Similarly, for
high $f$, the primary source of test error is the false negative rate $1 - \mathrm{TPR}$, which is minimized under the $\mathrm{OR}$ aggregation as per
in Eq.~\eqref{eq:inequalities}.

So far, we have shown that when tests with discriminatory power are used for diseases with prevalence $f < f_{\rm c}$, 
the $\mathrm{AND}$
aggregation protocol leads to the greatest reduction in the number of 
required tests when applied in series. Additionally, the $\mathrm{AND}$
aggregation protocol leads to larger PPV values compared
to the $\mathrm{OR}$ protocol.  Conversely, the
potential savings under the $\mathrm{OR}$ aggregation protocol are smaller, 
and the NPV is larger than under the $\mathrm{AND}$ protocol.
Thus, our analysis suggests that for $n=2$ tests, the most suitable protocol for minimizing test usage and maximizing the PPV estimate in low-prevalence scenarios is the series $\rm AND$ method.

To provide further analytical insight into the properties of repeated
tests, we consider aggregation functions involving more than two tests
in the next section.
\subsection*{Aggregating more than two tests}
\begin{table}
\footnotesize
\centering
\begin{tabular}{m{2cm} m{6cm} m{6cm}}\toprule
%\begin{tabular}{S{m{3.8cm}} S{m{5.5cm}} S{m{5.5cm}}}\toprule
    & \multicolumn{1}{c}{\textbf{parallel}} & \multicolumn{1}{c}{\textbf{series}} \\ \hline
\multicolumn{1}{c}{$\mathbf{AND}$} & 
\makecell[l]{
$n=2$: Slovenia (nationwide)~\cite{poljak2021seroprevalence}\\
} &  
\makecell[l]{
$n=2$: Norrbotten County, Sweden~\cite{seroprevalence-norrbotten}\\
$n=3$: Vienna, Austria~\cite{breyer2021low}
}
\\ \hline
\multicolumn{1}{c}{$\mathbf{OR}$}  & 
\makecell[l]{
$n=2$: South Africa (three communities) \cite{10.1093/cid/ciac198}\\
$n=4$: Santiago, Chile \cite{VIAL2022100606}
} & \multicolumn{1}{c}{-}
\\ \bottomrule
\end{tabular}
\vspace{1mm}
\caption{Examples of parallel and series test protocols that have been
  used in COVID-19 seroprevalence studies.}
\label{tab:examples}
\end{table}
In Table~\ref{tab:examples}, we list examples of SARS-CoV-2 seroprevalence studies 
where up to four tests were administered using various combinations of parallel and series ordering with
$\mathrm{AND}$ and $\mathrm{OR}$ aggregation~\cite{VIAL2022100606}. 
For $n=3$ tests, there are $r=2^3=8$ possible output sequences and $m= 2^{r}=2^8=256$ possible input-output mappings. 
For $n=4$, these numbers increase to $r=2^4=16$ and $m = 2^{r}=2^{16}=65,536$ respectively. 
Given the large number of possible ways
of combining $n$ tests, we will derive sensitivities and specificities
for a few select choices and otherwise resort to an algorithmic
evaluation of test performances as detailed in the following section.

Equations \eqref{eq:tpr12_par_and}--\eqref{eq:tpr12_tnr12_ser_and}
  show that for $n=2$, parallel and series test protocols carry the
  same sensitivities and specificities. This equivalence remains valid
  for $n\geq 3$ tests, so for notational simplicity we suppress the
  ``${\rm s}$'' and ``${\rm p}$'' superscripts that distinguish them.

For $n=3$ tests and an ${\rm AND}$ aggregation, the sensitivity and
specificity are
\begin{equation}
    {\rm TPR}_{1\land 2\land 3}={\rm TPR}_{1} {\rm TPR}_{2} {\rm TPR}_{3}
    \label{eq:n3_and_tpr}
\end{equation}
and
\begin{align}
\begin{split}
  {\rm TNR}_{1 \land 2 \land 3} = & {\rm TNR}_{1} + {\rm TNR}_{2}+{\rm TNR}_{3}
  -{\rm TNR}_{1}{\rm TNR}_{2}\\
  \: & \qquad  -{\rm TNR}_{1}{\rm TNR}_{3}
  -{\rm TNR}_{2}{\rm TNR}_{3}+{\rm TNR}_{1} {\rm TNR}_{2} {\rm TNR}_{3}\,,
\end{split}
\label{eq:n3_and_tnr}
\end{align}
respectively. Similarly, the sensitivity and specificity of an $\rm
OR$ test protocol with $n=3$ tests are
\begin{equation}
    \begin{split}
      {\rm TPR}_{1\lor 2\lor 3} = & {\rm TPR}_{1} + {\rm TPR}_{2}+{\rm TPR}_{3}
      -{\rm TPR}_{1}{\rm TPR}_{2}\\
      & \qquad -{\rm TPR}_{1}{\rm TPR}_{3}
      -{\rm TPR}_{2}{\rm TPR}_{3}+{\rm TPR}_{1} {\rm TPR}_{2} {\rm TPR}_{3}
    \end{split}
    \label{eq:n3_or_tpr}
\end{equation}
and
\begin{align}
\begin{split}
    {\rm TNR}_{1 \lor 2 \lor 3} & ={\rm TNR}_{1} {\rm TNR}_{2} {\rm TNR}_{3} \,.
\end{split}
\label{eq:n3_or_tnr}
\end{align}
The overall sensitivity and specificity of the
limiting $\mathrm{AND}$ and  $\mathrm{OR}$ aggregations for general $n$-tests are
\begin{equation}
  {\rm TPR}^{\mathrm{AND}}_{n} =  \prod_{i=1}^{n} {\rm TPR}_{i}\,, \quad
  {\rm TNR}^{\mathrm{AND}}_{n} =  1- \prod_{i=1}^{n} \big(1- {\rm TNR}_{i}\big)\,,
\end{equation}
and
\begin{equation}
  {\rm TPR}^{\mathrm{OR}}_{n} =   1- \prod_{i=1}^{n} \big(1- {\rm TPR}_{i}\big)\,,\quad 
  {\rm TNR}^{\mathrm{OR}}_{n} =  \prod_{i=1}^{n} {\rm TNR}_{i}\,.
\end{equation}
In line with Eq.~\eqref{eq:inequalities}, the TPRs and TNRs of the combined testing protocols
satisfy 
\begin{equation}
  {\rm TPR}^{\mathrm{AND}}_{n}  \leq {\rm TPR}_{i} \leq   {\rm TPR}^{\mathrm{OR}}_{n} \quad \mbox{and}
  \quad {\rm TNR}^{\mathrm{OR}}_{n} \leq {\rm TNR}_{i} \leq {\rm TNR}^{\mathrm{AND}}_{n} \quad \forall i \in \{1,\dots,n\} \,.
    \label{eq:inequalities_2}
\end{equation}
For odd $n\geq 3$, one can also employ a majority aggregation,
where at least $(n+1)/2$ tests have to be positive for the combined
test to be positive. The majority aggregation is intermediate
  relative to the ``all" and ``any"  characteristics of the ${\rm AND}$ and
${\rm OR}$ aggregations, respectively. The sensitivity of a majority
aggregation of $n=3$ tests is
\begin{equation}
    {\rm TPR}_{\mathrm{M}(1,2,3)}= {\rm TPR}_1 {\rm TPR}_2 + {\rm
      TPR}_1 {\rm TPR}_3 + {\rm TPR}_2 {\rm TPR}_3 - 2 {\rm TPR}_1
    {\rm TPR}_2 {\rm TPR}_3\,,
    \label{eq:n3_maj_tpr}
\end{equation}
and the corresponding specificity is
\begin{equation}
    {\rm TNR}_{\mathrm{M}(1,2,3)}= {\rm TNR}_1 {\rm TNR}_2 + {\rm
      TNR}_1 {\rm TNR}_3 + {\rm TNR}_2 {\rm TNR}_3 - 2 {\rm TNR}_1
    {\rm TNR}_2 {\rm TNR}_3\,.
    \label{eq:n3_maj_tnr}
\end{equation}
\begin{figure}[t!]
    \centering
    \includegraphics[width=\textwidth]{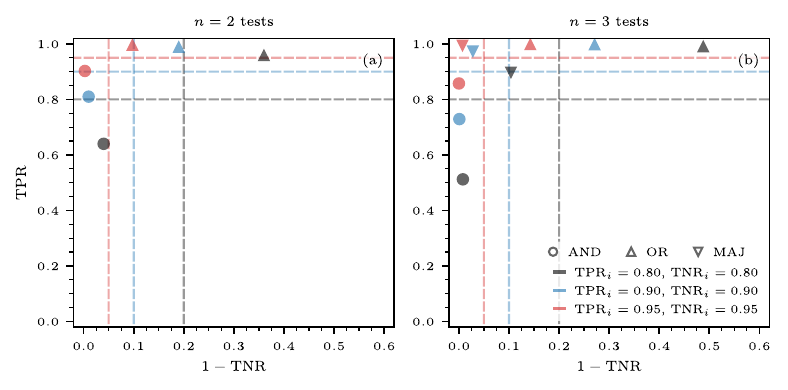}
    \caption{Receiver operating characteristic (ROC) curves for
      various combinations of tests and aggregation functions. (a) We
      consider $n=2$ tests and two distinct aggregation functions
      (disks: $\mathrm{AND}$ aggregation; triangles: $\mathrm{OR}$
      aggregation). (b) We consider $n=3$ tests and the same
      aggregation functions as in panel (a) along with the majority
      function represented by inverted triangles. Markers in black,
      blue, and red correspond to underlying tests $i\in\{1,\dots,n\}$ with sensitivities (${\rm TPR}_i$)
      and specificities (${\rm TNR}_i$) set to 0.8, 0.9, and 0.95,
      respectively. Dashed lines indicate the sensitivities and false
      positive rates (\ie, $1-\mathrm{TNR}$) of the individual
      isolated tests. Under $\mathrm{AND}$ aggregation, both the
      sensitivities and false positive rates of the aggregated tests
      are smaller than those of the individual tests. The opposite
      holds for $\mathrm{OR}$ aggregation. When considering $n=3$
      tests, the majority function results in higher sensitivities and
      smaller false positive rates compared to the individual isolated
      tests. This function provides a tradeoff between the 
      ``all'' and ``any'' characteristics of $\mathrm{AND}$ and $\mathrm{OR}$
      aggregations. The results shown are independent of the
        ordering (parallel or series) method used.}
    \label{fig:roc_n2_n3}
\end{figure}

In Figure~\ref{fig:roc_n2_n3}, we show receiver operating characteristic (ROC) curves for various combinations of tests and aggregation
functions. In Figure~\ref{fig:roc_n2_n3}(a), we present the
sensitivities and false positive rates for $\mathrm{AND}$
and $\mathrm{OR}$ aggregations with $n=2$ tests. Additionally, in
Figure~\ref{fig:roc_n2_n3}(b), we consider $\mathrm{AND}$
and $\mathrm{OR}$, and majority aggregation for
$n=3$ tests. Under $\mathrm{AND}$ aggregation, the sensitivities and
false positive rates of the aggregated tests are lower than those of
the individual tests. The opposite holds for $\mathrm{OR}$ aggregation. These findings
are in agreement with the analytical results in Eq.~\eqref{eq:inequalities_2}. 
Finally, when examining $n=3$ tests, the majority function yields
greater sensitivities and reduced false positive rates compared to the
individual isolated tests.
\subsection*{Efficiently combining $n$ tests}
For a given set of $n$ tests, what are the best aggregation protocols
in terms of sensitivities and specificities of the aggregated tests?
As discussed in the prior sections, there exist numerous possibilities
to combine individual tests, and the mathematical expressions for
aggregated sensitivities and specificities can be quite
lengthy. Therefore, we adopt an algorithmic approach to compute ROC
curves associated with $n$ tests, each potentially having distinct
sensitivities and specificities. In this context, we use the term ``efficient test'' to denote a test whose underlying sensitivity-specificity pairs are optimized to achieve the highest possible values.
\subsubsection*{Aggregation and optimization}
Algorithm 1 shown below computes the most efficient combination of $n$ conditionally independent tests
for given $\mathrm{TPR}_i$ and $\mathrm{TNR}_i$ of each test $i \in \{1,\dots,n\}$.
The following example illustrates our algorithm. We define
  $\mathcal{P}=\{P_1,\dots,P_{r}\}$ as the set of possible ordered outcomes deriving from the administration of 
  $n$ tests, where $r = 2^n$. For example, for $n=2$ tests, there are
  $r=2^n=2^2=4$ permutations and $P_1=(+,+)$, $P_2=(+,-)$, $P_3=(-,+)$,
  and $P_4=(-,-)$. Thus, we have
 \begin{equation}
      \mathcal{P}=\{(+,+),(+,-),(-,+),(-,-)\}\,.
  \end{equation}
  Each of these four outcomes can be mapped to either a positive aggregate diagnosis ``$+$'' or
  a negative one ``$-$''. Hence, there are $m=2^{r}=2^4=16$ mappings in total. For example, the output sequence 
$\mathcal{S}=(+,-,-,-)$ means that only the input $P_{1} =(+,+)$ is mapped to an aggregated
 ``$+$'', and the
other permutations $P_{2} =(+,-)$, $P_{3} =(-,+)$, and $P_{4} =(-,-)$ are
mapped to ``$-$.". This case corresponds to the 
$\mathrm{AND}$ aggregation protocol.  Similarly, 
$\mathcal{S}=(+,+,+,-)$ corresponds to the $\mathrm{OR}$ aggregation protocol.

We define the sensitivity $\mathrm{TPR}_\mathcal{S}$ associated with the
output sequence $\mathcal{S}=(S_1,\dots,S_r)$ in two steps. First,
we define $\mathrm{TPR}_\mathcal{S}$  as the sum over the
sensitivities $\mathrm{TPR}_{S_j}$ ($j\in\{1,\dots,r\}$) associated
with elements $S_j$ of $\mathcal{S}$. That is,
\begin{equation}
    \mathrm{TPR}_\mathcal{S} = \sum_{j=1}^{r} \mathrm{TPR}_{S_j}\,.
    \label{eq:n_test_sensitivity}
\end{equation}
Second, we define $\mathrm{TPR}_{S_j}$ as follows.
If element $S_j$ is ``$-$'' (\ie, if the input state $P_j$ gets
classified as negative), then we pose $\mathrm{TPR}_{S_j}=0$. Otherwise, if
element $S_j$ is ``$+$'', we calculate products of
$\mathrm{TPR}_i$ and $1-\mathrm{TPR}_i$ depending on whether the result from test
$i\in\{1,\dots,n\}$ is positive or negative. That is,
\begin{equation}
    \mathrm{TPR}_{S_j}=
    \begin{cases}
      \prod_{i=1}^n \left[ \mathrm{TPR}_i\delta_{i,+}
  +(1-\mathrm{TPR}_i)\delta_{i,-} \right] \,,\quad&\text{if~$S_j$~is~$+$}\\
        0\,,\quad&\text{if~$S_j$~is~$-$}
    \end{cases}\,,
    \label{eq:n_test_sub_sensitivity}
\end{equation}
where $\delta_{i,+}=1$ if test $i$ is positive and 0
otherwise. Likewise, $\delta_{i,-}=1$ if test $i$ is negative and
0 otherwise. For $n=2$ tests and $\mathcal{S}=(+,-,-,-)$, Eq.~\eqref{eq:n_test_sub_sensitivity} reduces to 
the $\rm AND$ aggregation $\mathrm{TPR}_{1 \land 2}$ result given in Eqs.~\eqref{eq:tpr12_par_and} 
and \eqref{eq:tpr12_tnr12_ser_and}. Similarly, for $n=2$ tests and 
$\mathcal{S}=(+,+,+,-)$, Eq.~\eqref{eq:n_test_sub_sensitivity} reduces to 
the $\rm OR$ aggregation $\mathrm{TPR}_{1 \lor2}$ result given in Eqs.~\eqref{eq:tpr12_par_or} 
and \eqref{eq:tpr12_tnr12_ser_or}. 
We follow the same steps to define the specificity 
$ \mathrm{TNR}_\mathcal{S}$ of the output sequence 
$\mathcal{S}$ so that 
\begin{equation}
    \mathrm{TNR}_\mathcal{S} = \sum_{j=1}^{r} \mathrm{TNR}_{S_j}
    \label{eq:n_test_specificity}
\end{equation}
where
\begin{equation}
    \mathrm{TNR}_{S_j}=
    \begin{cases}
        0\,,\quad&\text{if~$S_j$~is~$+$}\\
        \prod_{i=1}^n \left[(1-\mathrm{TNR}_i)\delta_{i,+}+
        \mathrm{TNR}_i\delta_{i,-} \right]\,,\quad&\text{if~$S_j$~is~$-$}
    \end{cases}\,.
    \label{eq:n_test_sub_specificity}
\end{equation}
For $n=2$ tests and $\mathcal{S}=(+,-,-,-)$, Eq.~\eqref{eq:n_test_sub_specificity} reduces to 
the $\rm AND$ aggregation $\mathrm{TNR}_{1 \land 2}$ result given in Eqs.~\eqref{eq:tnr12_par_and} 
and \eqref{eq:tpr12_tnr12_ser_and}. Similarly, for $n=2$ tests and $\mathcal{S}=(+,+,+,-)$, Eq.~\eqref{eq:n_test_sub_specificity} simplifies to 
the $\rm OR$ aggregation $\mathrm{TNR}_{1 \lor2}$ result given in Eqs.~\eqref{eq:tnr12_par_or} 
and \eqref{eq:tpr12_tnr12_ser_or}. 

We identify two limit cases. One is the output sequence $\mathcal{S}=(+,+,+,+)$ where all
input permutations $P_{j}$ are mapped to ``$+$" outcomes. In this case, the 
aggregated sensitivity and specificity are $\mathrm{TPR}_\mathcal{S}=1$ and
$\mathrm{TNR}_\mathcal{S}=0$, respectively. The other limit case
 is the output sequence $\mathcal{S}=(-,-,-,-)$ where all
input permutations $P_{j}$ are mapped to ``$-$" outcomes. Here,
the aggregated sensitivity and specificity are $\mathrm{TPR}_\mathcal{S}=0$ and
$\mathrm{TNR}_\mathcal{S}=1$, respectively.

Once we have determined all pairs
$(\mathrm{TPR}_\mathcal{S},\mathrm{TNR}_\mathcal{S})$ associated with
the $m=2^{r}$ test aggregations, we identify the most efficient test combinations, \ie, those combinations where the underlying sensitivity-specificity pairs reach the highest values. This is achieved by employing a convex-hull algorithm, such as Graham scan~\cite{Graham72} and
Quickhull~\cite{greenfield1990,barber1996quickhull}, to determine the ROC frontier in the 
$(\mathrm{TPR}_\mathcal{S},1-\mathrm{TNR}_\mathcal{S})$ space (\ie, true
positive-false positive rate space). We summarize all steps of our
algorithm in \texttt{Python} pseudocode in
Algorithm~\ref{alg:n_test_aggregation}.
\begin{algorithm}
\footnotesize
\caption{Compute the most efficient combinations of $n$ conditionally independent tests.
}
\label{alg:n_test_aggregation}
\begin{algorithmic}[1]
\Inputs{$n$, $\mathrm{TPRs}$, $\mathrm{TNRs}$, $\Call{ConvexHull}$}
\Outputs{roc\_frontier}
\State input\_permutations $\gets$ list(itertools.product([0, 1], repeat=$n$)) 
\Comment{\,Generate input permutations $\mathcal{P}$}
\State input\_output\_mappings $\gets$ list(itertools.product([0, 1], repeat=$2^n$))
\Comment{\,Generate output sequences $\mathcal{S}$}
\State TPR\_arr $\gets$ []
    \State TNR\_arr $\gets$ []
    \For{input\_output\_map in input\_output\_mappings}
        \State TPR\_combined $\gets$ []
        \State TNR\_combined $\gets$ []
        \For{perm, output\_value \textbf{in} zip(input\_permutations, input\_output\_map)}
            \If{output\_value}
                \State TPR\_combined.append($\prod_{i=0}^{n-1}$ (TPRs[i] \textbf{if} perm[i] \textbf{else} 1-TPRs[i]))
                \Comment{\,see Eq.~\eqref{eq:n_test_sub_sensitivity}}
            \Else
                \State TNR\_combined.append($\prod_{i=0}^{n-1}$ (1-TNRs[i] \textbf{if} perm[i] \textbf{else} TNRs[i]))
                \Comment{\,see Eq.~\eqref{eq:n_test_sub_specificity}}
            \EndIf
        \EndFor
        \State TPR\_arr.append(sum(TPR\_combined)) \Comment{\,Compute aggregated sensitivity using Eq.~\eqref{eq:n_test_sensitivity}}
        \State TNR\_arr.append(sum(TNR\_combined))
        \Comment{\,Compute aggregated specificity using Eq.~\eqref{eq:n_test_specificity}}
    \EndFor
    \State points $\gets$ concatenate(1-TNR\_arr, TPR\_arr)
    \State convex\_hull $\gets$ \Call{ConvexHull}{points}
    \State roc\_frontier $\gets$ []
    \For{edge \textbf{in} convex\_hull}
        \If{(points[edge, 1][0] $\geq$ points[edge, 0][0]) \textbf{and} (points[edge, 1][1] $\geq$ points[edge, 0][1])}
            \State roc\_frontier.append(points[edge])
            \Comment{\,ROC points must satisfy $\mathrm{TPR}\geq 1-\mathrm{TNR}$}
        \EndIf
    \EndFor
\State \textbf{return} roc\_frontier
\end{algorithmic}
\end{algorithm}
\subsubsection*{An example with three antigen tests}
As an example, we apply Algorithm 1 to the aggregation of $n=3$ commonly used
SARS-CoV-2 antigen tests~\cite{dinnes2022rapid}. We list their median
sensitivities and specificities along with their 95\% confidence
intervals (CIs) in Table~\ref{tab:tests}. In this case there are $r=2^n = 2^3= 8$
permutations of test results and $m=2^r = 256$ possible input-output mappings.
The set of permutations is 
\begin{equation}
\label{perm3}
{\mathcal P} = \{(+,+,+),(+,+,-), (+,-,+), (+,-,-), (-,+,+), (-,+,-), (-,-,+), (-,-,-) \}\,,
\end{equation}
and the corresponding output sequence is ${\mathcal S} = (S_1, \dots, S_r)$, where $r = 8$ and $S_j \in\{+,-\}$.  To make the notation simpler, we introduce the Boolean variable $Y_i \in \{0,1\}$
for each test $i \in \{1,\dots,n\}$ and map the aggregation method
$\mathcal S$ to its corresponding Boolean expression as shown below.

\begin{figure}[h]
    \centering
    \includegraphics[width=\textwidth]{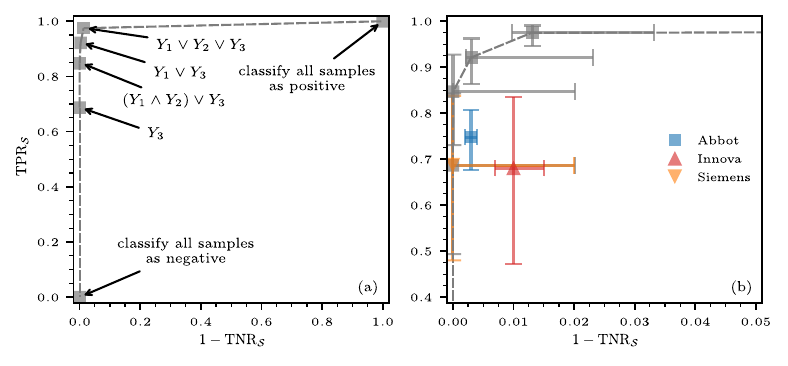}
    \caption{ROC curves associated with the aggregation of three
      antigen tests (Abbot, Innova, and Siemens). The
      sensitivities and specificities of the $n=3$ tests are listed in 
      Table~\ref{tab:tests}. (a) The ROC curve associated with the
      aggregation of the three antigen tests as derived from 
      Eqs.\,\eqref{eq:n_test_sub_sensitivity} and  \eqref{eq:n_test_sub_specificity}.         
          We use $Y_i\in\{0,1\}$
      to denote the outcome of test $i\in\{1,2,3\}$. (b) A magnified
      view of the ROC curve without the trivial aggregated tests
      that classify all samples as either negative or positive. The
      error bars indicate the 95\% CIs that we generated from $10^6$
      samples of beta distributions capturing the 95\% CIs of the
      underlying individual sensitivities and specificities.}
    \label{fig:roc_antigen}
\end{figure}

We begin by using the median sensitivities and specificities of the
three tests from Table~\ref{tab:tests} as inputs in Algorithm~\ref{alg:n_test_aggregation} for various
aggregation protocols and use them to derive the 
corresponding ROC curve shown in
Figure~\ref{fig:roc_antigen}(a). On this curve, there exist two
limit cases: (i) an aggregation method where both
sensitivity $\mathrm{TPR}_{\mathcal S}$ and false positive rate 
$1-\mathrm{TNR}_{\mathcal S}$
are equal to 0, effectively classifying all input sequences as negative. This corresponds to 
$S_j = -$  for all $j \in \{1,\dots,8\}$;  and (ii) an aggregation
method with sensitivity $\mathrm{TPR}_{\mathcal S}$ and false positive rate
$1 - \mathrm{TNR}_{\mathcal S}$
  both at 1, resulting in the
classification of all input sequences as positive, corresponding to $S_j = +$
for all $j \in \{1,\dots,8\}$.

We also include four more aggregation methods
on the ROC curve shown in Figure~\ref{fig:roc_antigen}(a).
The first of these requires that only the last test (\ie, Siemens) be positive,
irrespective of the outcomes of the other two. This aggregation method
corresponds to ${\mathcal S} = (+,-,+,-,+,-,+,-)$ and is denoted $Y_3$. 
It exhibits the smallest possible false positive rate,   $1 - \mathrm{TNR}_{\mathcal S} = 0$, which is intuitive given that the Siemens test also has the lowest median false positive
rate of 0. Its sensitivity is $\mathrm{TPR}_{\mathcal S} =
68.7\%. $

\begin{table}[t!]
\renewcommand*{\arraystretch}{2.0}
\footnotesize
\centering
\begin{tabular}{m{7.5cm} m{4cm} m{4cm}} \toprule
% \begin{tabular}{S{m{2.8cm}} S{m{4cm}} S{m{4cm}}} \toprule
    & \multicolumn{1}{c}{\textbf{sensitivity}} & \multicolumn{1}{c}{\textbf{specificity}} \\ \hline
\textbf{Abbott – Panbio COVID-19 Ag} & \multicolumn{1}{c}{74.8\% (67.6 -- 80.8\%)} & \multicolumn{1}{c}{99.7\% (99.6 -- 99.8\%)} \\ \hline
\textbf{Innova Medical Group - Innova SARS-CoV-2 Ag} & \multicolumn{1}{c}{68.1\% (47.2 -- 83.6\%)} & \multicolumn{1}{c}{99.0\% (98.5 -- 99.3\%)} \\ \hline
\textbf{Siemens - CLINITEST Rapid COVID-19 Ag} & \multicolumn{1}{c}{68.7\% (48.0 -- 83.8\%)} & \multicolumn{1}{c}{100\% (98.0 -- 100\%)} \\ \bottomrule
\end{tabular}
\vspace{1mm}
\caption{Median sensitivities and specificities of three commonly used
  SARS-CoV-2 antigen tests that are based on studies involving
  symptomatic patients~\cite{dinnes2022rapid}. Numbers in parentheses
  denote 95\% CIs.}
\label{tab:tests}
\end{table}

The subsequent aggregated test result shown on the ROC curve requires both the first and the second tests (\ie, Abbott and Innova), or
the last one (\ie, Siemens) to yield positive results. This protocol corresponds to 
$\mathcal{S} = (+,+,+,-,+,-,+,-)$ and can be written in Boolean algebra as 
$(Y_1\land Y_2)\lor Y_3$. Using the values listed in Table \ref{tab:tests}
and Eqs.~\eqref{eq:n_test_sub_sensitivity} and \eqref{eq:n_test_sub_specificity}, it can be verified that its sensitivity and false positive rate are
$\mathrm{TPR}_{\mathcal S}=  84.6\%$ and $1 - \mathrm{TNR}_{\mathcal S} = 3.0 \times 10^{-3}\%$, respectively.

We can improve the aggregated sensitivity 
by omitting the second test (\ie, Innova), which has the lowest sensitivity at 68.1\%; 
the tradeoff is to accept a slightly higher false positive rate. 
This protocol yields the next point on the ROC curve. It corresponds to $\mathcal S = \{+, +, +, +,+, -,+,- \}$
and can be written using an $\mathrm{OR}$ aggregation over the first and last tests,  
 \ie, $Y_1\lor Y_3$. Equations~\eqref{eq:n_test_sub_sensitivity} and \eqref{eq:n_test_sub_specificity}
yield the sensitivity $\mathrm{TPR}_{\mathcal S}=  92.1 \%$
and the false positive rate  $ 1 -  \mathrm{TNR}_{\mathcal S} = 0.3\%$.

Finally, the largest sensitivity smaller than 100\% is achieved 
through an $\mathrm{OR}$ aggregation over all tests 
\ie, for $Y_1\lor Y_2 \lor Y_3$. This corresponds to $\mathcal{S} = \{+,+,+,+,+,+.+,-\} $, an output sequence
with sensitivity $\mathrm{TPR}_{\mathcal S}=  97.5\%$ and false
positive rate $ 1 -  \mathrm{TNR}_{\mathcal S} = 1.3\%$ as per
Eqs.~\eqref{eq:n_test_sub_sensitivity} and \eqref{eq:n_test_sub_specificity}.

For a more detailed comparison between aggregated and individual tests, we show a magnified view of the four non-trivial aggregations in the ROC frontier in Figure~\ref{fig:roc_antigen}(b) and include individual tests.  In this plot, we
incorporate CIs alongside median sensitivities and false positive
rates. We generate these CIs from $10^5$ samples of beta distributions
capturing the 95\% CIs of the underlying individual sensitivities and
specificities (see Materials and methods for further details). We
observe that the two $\mathrm{OR}$ protocols, $Y_1 \lor Y_3$ and $Y_1
\lor Y_2 \lor Y_3$, exhibit significantly higher sensitivity compared
to each individual test.
\subsection*{Estimating prevalence}
In the preceding sections, we have observed how repeating and
aggregating results from diagnostic and screening tests can substantially enhance
sensitivity and specificity. This enhancement can contribute to
improved infectious-disease surveillance and
management~\cite{bottcher2021using,zhang2022data,schneider2022epidemic,bottcher2022statistical}
by providing more accurate estimates $\hat{f}$ of the true prevalence
$f$ in a population. The prevalences $\hat{f}$ and $f$ may
be time-dependent and stratified, \textit{e.g.}, according to
age, (\ie, $\hat{f}\equiv \hat{f}(a_k,t)$ and $f\equiv f(a_k,t)$ where $a_k$ is a given age). 
\subsubsection*{Correcting test errors}
We distinguish between the measured prevalence $\hat{f}_{\mathcal{S}}^*(a_k,t)$
of a disease derived from testing a given sample of the population and using the
aggregation method $\mathcal S$, 
and the measured, error-corrected prevalence $\hat{f}(a_k,t)$, 
which is instead the resulting estimate of the true disease prevalence $f(a_k, t)$.  
If we also assume that the selected sample is unbiased and 
representative of the infection behavior in the entire population,  we can 
identify the estimate $\hat f(a_k,t)$ with the actual prevalence $f(a_k, t)$ 
and write $\hat{f}(a_k,t) = f(a_k, t).$
For $n$ aggregated tests with output
sequence $\mathcal{S}$, the quantities
$\hat{f}_{\mathcal{S}}^*(a_k,t)$ and $\hat{f}(a_k,t)$ are related via
\begin{equation}
\label{eq:fs}
  \hat{f}_{\mathcal{S}}^*(a_k,t)=\hat{f}(a_k,t)\mathrm{TPR}_\mathcal{S}
  +(1-\hat{f}(a_k,t))(1-\mathrm{TNR}_\mathcal{S})\,,
\end{equation}
which yields
\begin{equation}
  \hat{f}(a_k,t) = \frac{\hat{f}_{\mathcal{S}}^*(a_k,t)
    +\mathrm{TNR}_\mathcal{S}-1}{\mathrm{TPR}_\mathcal{S}+\mathrm{TNR}_\mathcal{S}-1}\,,
    \label{eq:f_hat}
\end{equation}
a generalized Rogan--Gladen prevalence
estimate~\cite{rogan1978estimating} that accounts for the sensitivity
and specificity of the aggregated tests with output sequence
$\mathcal{S}$. We omit the subscript $\mathcal{S}$ in
  $\hat{f}(a_k,t)$ since the error-corrected
  prevalence is an estimate of the true prevalence and should not depend of the method used for aggregating
  test results.
For example, for $n=2$ tests under
$\mathrm{AND}$ and $\mathrm{OR}$ aggregation and using
  Eqs.~\eqref{eq:tpr12_par_and}--\eqref{eq:tpr12_tnr12_ser_and}, we
have
\begin{equation}
  \hat{f}(a_k,t)=\frac{\hat{f}_{1\land 2}^*(a_k,t)
    +{\rm TNR}_{1}+{\rm TNR}_{2}
    -{\rm TNR}_{1}{\rm TNR}_{2}-1}{{\rm TPR}_{1} {\rm TPR}_{2}+{\rm TNR}_{1}
    +{\rm TNR}_{2}-{\rm TNR}_{1}{\rm TNR}_{2}-1}, 
    \label{eq:f_hat_and}
\end{equation}
and
\begin{equation}
  \hat{f}(a_k,t)=\frac{\hat{f}_{1\lor 2}^*(a_k,t)+
    {\rm TNR}_{1}{\rm TNR}_{2}-1}{{\rm TPR}_{1}+{\rm TPR}_{2}
    -{\rm TPR}_{1} {\rm TPR}_{2}+{\rm TNR}_{1}{\rm TNR}_{2}-1}\,,
\end{equation}
respectively.
\begin{figure}[t!]
    \centering
    \includegraphics[width=\textwidth]{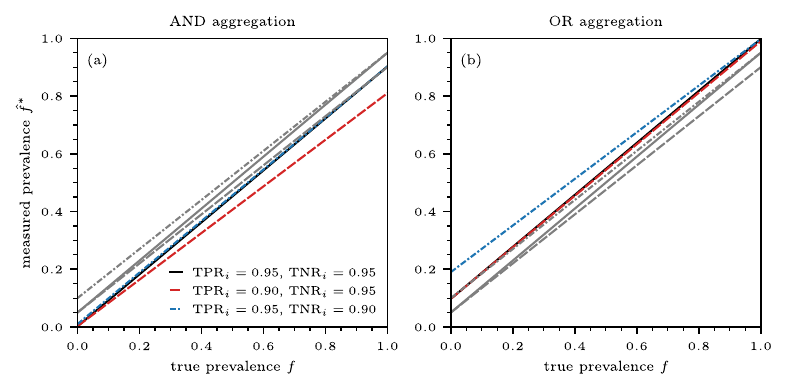}
    \caption{Measured prevalence $\hat{f}^*$ as a function of true
      prevalence $f$ under the assumption that the measured, error-corrected prevalence
      $\hat f$ in Eq.~\eqref{eq:fs} can be identified with the true prevalence $f$. 
       The results that we show in panels (a) and (b)
      are based on $\mathrm{AND}$ and $\mathrm{OR}$ aggregations of
      two tests $i\in\{1,2\}$, respectively. We consider three
      different combinations of true positive and true negative rates
      (solid black lines: $\mathrm{TPR}_i=0.95$ and
      $\mathrm{TNR}_i=0.95$; dashed red lines: $\mathrm{TPR}_i=0.90$
      and $\mathrm{TNR}_i=0.95$; dash-dotted blue lines:
      $\mathrm{TPR}_i=0.95$ and $\mathrm{TNR}_i=0.90$). Grey lines
      indicate measured prevalences associated with individual tests.}
    \label{fig:prevalence_and_or}
\end{figure}

In Figure~\ref{fig:prevalence_and_or}, we show the measured
(uncorrected) prevalences $\hat{f}^*_{1 \land 2}$ 
and $\hat{f}^*_{1 \lor 2}$
associated with the $\mathrm{AND}$
and $\mathrm{OR}$ aggregations 
using Eq.\,\eqref{eq:fs} and 
Eqs.~\eqref{eq:tpr12_par_and}--\eqref{eq:tpr12_tnr12_ser_and} for 
the corresponding  $\mathrm{TPR}_{\cal S}$ and  $\mathrm{TNR}_{\cal S}$, 
and using different sensitivities and specificities for $n=2$ tests. 
For simplicity, we assume that samples are unbiased and that
the measured, error-corrected prevalence $\hat f$ can be identified 
with the true prevalence $f$. 

 In line with our findings regarding PPV and NPV and the trends shown in 
 Figure~\ref{fig:ppv_vs_npv}, we observe in Figure~\ref{fig:prevalence_and_or} that $\hat{f}^*_{1 \land 2}$, the measured prevalence  under
$\mathrm{AND}$ aggregation, deviates only slightly from the
true prevalence $f$, for low true prevalence, whereas under $\mathrm{OR}$ aggregation
deviations between $\hat{f}^*_{1 \lor 2}$ and $f$ are small at large true prevalence.

As a real-world example of prevalence correction under aggregated
testing, we consider the seroprevalence study from Norrbotten Sweden
(May 25 -- June 5, 2020)~\cite{seroprevalence-norrbotten}. In this
study, two SARS-CoV-2 tests, Abbott SARS-CoV-2 IgG and the Euroimmun
Anti-SARS-CoV-2 ELISA (IgG), were administered to an age-stratified population
and combined using an
$\mathrm{AND}$ function. The non age-stratified  measured prevalence is estimated at
$\hat f^*_{1 \land 2} = 1.9\%$; other details of the testing protocols employed in this study
are discussed in the Materials and methods. We use the $\hat f^*_{1 \land 2}$ estimate to 
calculate the measured, error-corrected prevalence using Eq.~\eqref{eq:f_hat_and} and 
the underlying individual test sensitivities and specificities given in the Materials and methods section. 
We also calculate the corresponding 95\% CIs by generating $10^6$ samples from beta distributions capturing the
measured prevalence $\hat f^{*}_{1 \land 2}$. 
We present the age-stratified measured prevalences {$\hat f ^*_{1 \land 2}$} and the 
corresponding 
measured, error-corrected prevalences $ \hat f $  for various age groups in Table~\ref{tab:norrbotten}.

Equation~\eqref{eq:f_hat_and} yields a non-stratified measured, error-corrected
prevalence $\hat f =  2.5\%$ (1.1 -- 5.0\%), which is higher than the measured prevalence 
$\hat f^*_{1 \land 2} = 1.9\%$ (0.8 - 3.7\%).  Because the sensitivity of tests combined using an $\mathrm{AND}$ function is lower compared to the sensitivity of the underlying constituent tests, the measured prevalence associated with this aggregation function usually underestimates the true prevalence. Hence,
the measured, error-corrected prevalence is substantially larger in this example
than the measured one.
\begin{table}
\renewcommand*{\arraystretch}{2.5}
\footnotesize
\centering
\begin{tabular}{S{m{4cm}} S{m{4cm}} S{m{4cm}}} \toprule
\multicolumn{1}{c}{\textbf{age group}} & \multicolumn{1}{c}{\textbf{measured prevalence}} $\hat{f}_{1\land 2}^*$ & \multicolumn{1}{c}{\textbf{error-corrected prevalence}} $\hat{f}$ \\ \hline
\multicolumn{1}{c}{20--29 years} & \multicolumn{1}{c}{6.6\% (1.8 -- 15.9\%)} & \multicolumn{1}{c}{8.8\% (2.4 -- 21.6\%)} \\ \hline
\multicolumn{1}{c}{30--64 years} & \multicolumn{1}{c}{0.7\% (0.1 -- 2.7\%)} & \multicolumn{1}{c}{0.9\% (0.1 -- 3.3\%)} \\ \hline
\multicolumn{1}{c}{65--80 years} & \multicolumn{1}{c}{2.1\% (0.3 -- 7.3\%)} & \multicolumn{1}{c}{2.8\% (0.4 -- 9.5\%)} \\ \bottomrule
\end{tabular}
\vspace{1mm}
\caption{Measured and error-corrected prevalence in Norrbotten, Sweden
  (May 25 -- June 5, 2020)~\cite{seroprevalence-norrbotten}. The error
  correction method we employed takes into account the two tests used
  in the seroprevalence study from Norrbotten: (i) the Abbott
  SARS-CoV-2 IgG kit and (ii) the Euroimmun Anti-SARS-CoV-2 ELISA
  (IgG). These tests have been aggregated using an $\mathrm{AND}$
  function. We calculated the measured, error-corrected prevalence
   through     Eq.~\eqref{eq:f_hat_and} and their corresponding 95\% CIs by
  generating $10^6$ samples from beta distributions capturing the
  measured prevalence as well as the underlying individual test
  sensitivities and specificities. Details of the study are listed in 
  Materials and methods.}
\label{tab:norrbotten}
\end{table}
\subsubsection*{An application in fatality and hospitalization monitoring}
Prevalence estimates commonly arise in infection fatality and
hospitalization ratios, which are useful measures for monitoring
outbreak severity. For a given jurisdiction at time $t$, the infection
fatality ratio ${\rm IFR}(a_k, t)$ of the population of age in the interval
$[a_k,a_{k+1})$ is
\begin{equation}
    {\rm IFR}(a_k,t)=\frac{D(a_k,t)}{f(a_k,t) N(a_k)}\,,
    \label{eq:crude_ifr}
\end{equation}
where $f(a_k,t)$ and $D(a_k,t)$ respectively denote the age-stratified
true proportion of infected individuals at time $t$ and the total
number of  infection-caused fatalities up to time $t$
measured from the start of an outbreak and within the age interval 
$[a_k,a_{k+1})$. In the above definition, we
assume that the overall population $N(a_k)$ of age in the interval
$[a_k,a_{k+1})$ is constant in the time horizon of interest. The
  denominator $f(a_k,t) N(a_k)$ in Eq.~\eqref{eq:crude_ifr} quantifies
  the total number of age-stratified infections at time $t$ since the
  start of an outbreak (\ie, current and prior infections).

The number of infection-caused fatalities, $D(a_k,t)$, may be
difficult to infer because of various confounding factors. These
factors include variations in protocols for attributing the cause of
death, the existence of co-morbidities~\cite{cdc2020}, and delays in
reporting. In jurisdictions where underreporting is prevalent,
statistics on excess deaths may offer a more accurate assessment of
the overall death
toll~\cite{bottcher2021using,bottcher2022statistical}.

Analogous to the IFR, the infection hospitalization ratio $\mathrm{IHR}(a_k, t)$ of the
population of age in the interval $[a_k,a_{k+1})$ in a given
  jurisdiction is
\begin{equation}
    {\rm IHR}(a_k,t)=\frac{H(a_k,t)}{f(a_k,t) N(a_k)}\,,
    \label{eq:crude_ihr}
\end{equation}
where $H(a_k,t)$ is the corresponding total number of age-stratified
infection-caused hospitalizations up to time $t$ measured from the
start of an outbreak. Because of the time lag between infection and
resolution, both the IFR and IHR may underestimate the true burden of
an outbreak, especially in the early stages when the number of new
cases increases rapidly~\cite{bottcher2020case}. In Table~\ref{tab:testing_variables}, we summarize the main variables used in outbreak severity measures.

The true proportion of infections $f(a_k,t)$ used in the denominators
of both $\mathrm{IFR}$ and $\mathrm{IHR}$ is usually difficult to
quantify for large populations. We can thus employ prevalence
estimates $\hat{f}(a_k,t)$ as derived in Eq.~\eqref{eq:f_hat} that are usually based
on serological testing of random samples of the entire population. Estimated proportions of
infections $\hat{f}(a_k,t)$ that have been obtained using serological tests can be assumed to be close to the true proportions $f(a_k,t)$ if
antibody waning is negligible and if the population sample is unbiased
and representative of the whole population.

We denote the corresponding IFR and IHR estimates by
\begin{equation}
    \widehat{{\rm IFR}}(a_k,t)=\frac{D(a_k,t)}{\hat{f}(a_k,t) N(a_k)}\,,
    \label{eq:test_adj_ifr}
\end{equation}
and
\begin{equation}
    \widehat{\rm IHR}(a_k,t)=\frac{H(a_k,t)}{\hat{f}(a_k,t) N(a_k)}\,,
    \label{eq:test_adj_ihr}
\end{equation}
respectively.
\begin{table}
\footnotesize
\centering
\renewcommand*{\arraystretch}{1.6}
\begin{tabular}{ >{\centering\arraybackslash} m{10em}>{\centering\arraybackslash} m{28em}}\toprule
\textbf{Symbol} & \textbf{Definition}
\\[1pt] \midrule
\,\,\, $N(a_k)\in \mathds{N}$\,\, & population of age in the interval
$[a_k,a_{k+1})$ in a given jurisdiction\\[1pt] \hline
\,\,\, $D(a_k,t)\in \mathds{N}$\,\, & total number of infection-caused
fatalities of age in the interval $[a_k,a_{k+1})$ in a given
  jurisdiction at time $t$ (measured from the start of an outbreak)
  \\[1pt] \hline
\,\,\, $H(a_k,t)\in \mathds{N}$\,\, & total number of infection-caused
hospitalizations of age in the interval $[a_k,a_{k+1})$ in a given
  jurisdiction at time $t$ (measured from the start of an outbreak)
  \\[1pt] \hline
\,\,\, $f(a_k,t): [0,1]$\,\, & true proportion of infected individuals
of age in the interval $[a_k,a_{k+1})$ at time $t$ in a given
  jurisdiction \\[1pt] \bottomrule
\end{tabular}
\vspace{1mm}
\caption{Main variables used in outbreak severity
  measures. Population, fatality, hospitalization, and prevalence
  statistics are often reported for $N_{a}$ age intervals
  $[a_{k-1},a_k)$ ($k\in\{1,\dots,N_a\}$) with $a_k=a_0+\sum_{\ell=1}^{k}
    \Delta a_\ell$. Here, $a_0$ is the smallest age value
    in the data set and $\Delta a_\ell$ is the width of the $\ell$-th age
    window. We assume that the population size $N(a_k)$ is constant in
    the considered time window. The closed interval $[0,1]$ contains
    0, 1, and all numbers in between, and $\mathds{N}$ denotes the set
    of non-negative integers.}
\label{tab:testing_variables}
\end{table}

The seroprevalence study from Norrbotten, Sweden (May 25
-- June 5, 2020)~\cite{seroprevalence-norrbotten} yields an overall measured, error-corrected seroprevalence 
of $\hat f = 2.5\%$ (1.1 -- 5.0\%). We assume that this prevalence estimate, obtained for a subpopulation aged 20 to 80 years, is reflective of the prevalence in the entire population of 249,614 individuals.
Using the total number of 59 fatalities and 242 hospitalizations documented throughout the entire study duration, along with Eqs.~\eqref{eq:test_adj_ifr}
and  \eqref{eq:test_adj_ihr}, we obtain
$\widehat{\mathrm{IFR}}=0.9\%$ (0.5 -- 2.2\%) and
$\widehat{\mathrm{IHR}}=3.8\%$ (1.9 -- 9.1\%). These values are lower
than the fatality ratio of 1.2\% (0.6 -- 3.0\%) and hospitalization
ratio of 5.1\% (2.6 -- 12.1\%) obtained with the uncorrected, measured
prevalence $\hat f^*_{1 \land 2} = 1.9\%$ (0.8 - 3.7\%).
\section*{Discussion}
Repeating and aggregating results from diagnostic and screening tests
can significantly enhance overall test performance. While our primary
focus has been on aggregating tests within the context of infectious-disease surveillance, similar concepts hold broad clinical
applicability, such as in diabetes
testing~\cite{brohall2006prevalence,kermani2017accuracy}, medical
imaging~\cite{weinstein2005clinical,zou2006statistical,brennan2019benefits},
and cancer
screening~\cite{neuhauser1975we,collins2005accuracy}. The
  complex clinical conditions are usually probed by tests performing
  multiclass discrimination, requiring generalizations of the ROC surface and other reduction schemes~\cite{multi_roc}.
 
Starting from the aggregation of two tests, for which there are 16
fundamental two-input Boolean gates, we derived expressions for
the sensitivity and specificity of aggregated tests, assuming their
conditional independence. We quantified the potential for saving
tests when employing series testing compared to parallel testing,
without compromising sensitivity and specificity. Additionally, we
discussed the strong dependence of the positive predictive
value (PPV) (\ie, the ratio of true positives to positive calls) and
negative predictive value (NPV) (\ie, the ratio of true negatives to
negative calls) on the employed aggregation mechanism. For example,
$\mathrm{AND}$ aggregation yields good relative large PPVs and NPVs at low prevalence
values, while $\mathrm{OR}$ aggregation does so for larger
prevalences.

Expressions of sensitivity and specificity for aggregations of more
than two tests can also be derived. However, these expressions may
become very lengthy. Thus, we developed an algorithm capable of
identifying the best way  
of aggregating results from a given set of tests in terms of
efficient sensitivity-specificity pairs (\ie, sensitivity-specificity
values that lie on an ROC frontier). We applied this algorithm to
three commonly used SARS-CoV-2 tests and demonstrated how their
individual sensitivities and specificities can be significantly
improved when combined.

Finally,  in this work we established a connection between aggregating tests
and prevalence estimates in infectious-disease surveillance. Such
estimates are pertinent for computing measures like the infection
fatality ratio (IFR) and infection hospitalization ratio (IHR).

Although our work addresses various factors related to aggregating tests, there are additional aspects that we have not considered. 
For instance, certain tests may entail higher costs or varying levels of complications for patients (see, \eg, chapter 7.4 in \cite{felder2022}). Other refinements may incorporate test-avoidance, or increasing levels of 
test-fatigue when multiple tests are to be administered. Incorporating these effects requires formulating appropriate target functions 
and adjusting our optimization approach. In the context of an ROC curve, a target function that quantifies the utility gain associated with treating a sick individual and the utility loss associated with treating a healthy individual enables the identification of the optimum aggregation approach~\cite{felder2003priori}.
Moreover, although we have incorporated sensitivity and specificity data for numerous tests in our analysis, it would be beneficial to validate the results of our model through experimental data on aggregated tests.

In addition to the described applications, our work can help inspire aggregation methods in social choice
theory and decision-making under uncertainty, where the objective is
to effectively combine individual
opinions~\cite{smoke1960reliable,brandt2016handbook,
  bottcher2022examining,kurvers2023automating,bottcher2024collective}. For instance, it can inform decision-making
processes in organizations where decision makers also possess
sensitivities and specificities with respect to a given decision
task. Furthermore, our paper is closely connected to historical works
on fault-tolerant computing by von
Neumann~\cite{von1956probabilistic}, Moore, and Shannon
\cite{moore1956a, moore1956b, kochen1959extension}, who studied how reliable (Boolean)
computing elements can be constructed from unreliable components.
\section*{Materials and methods}
\subsection*{Beta distribution sampler}
To calculate CIs associated with aggregated tests and related
quantities that depend on multiple factors such as sensitivity,
specificity, and prevalence, we employ a Monte Carlo sampling
technique. In this work, we consider samples drawn from a beta
distribution
\begin{equation}
  \mathbb{P}(x; \alpha, \beta) = \frac{\Gamma(\alpha)\Gamma(\beta)}{\Gamma(\alpha + \beta)}
  x^{\alpha-1}(1-x)^{\beta-1}\,,
\end{equation}
where $x\in[0,1]$, $\alpha,\beta$ are shape parameters, and
$\Gamma(\cdot)$ denotes the gamma function. Sensitivities,
specificities, and prevalences are quantities with a support of
$[0,1]$, so beta distributions are plausible approximations of their
underlying distributions.

We determine shape parameters such that the corresponding
distributions capture the median and 95\% CIs of the underlying
quantities. To do so, we minimize the sum of squared differences
between the cumulative distribution at the 2.5\%, 50\%, and 97.5\%
quantiles, and the corresponding empirical median and 95\% CI
values. We carry out this optimization process by employing the
\texttt{fmin} function implemented in \texttt{scipy.optimize} in \texttt{Python}. 
Further
implementation details are available at \cite{gitlab}.
\subsection*{Hospitalization, fatality, and serology data}
In the main text, we use data from a seroprevalence study conducted in
Norrbotten, Sweden, during weeks 22 and 23 of 2020 (May 25 to June
5)~\cite{seroprevalence-norrbotten}. We considered the
hospitalization, fatality, and seroprevalence data provided in this
study to illustrate how errors associated with aggregated tests can be
addressed. The study encompassed a population of 182,828 adults aged 20
to 80 years. The age distribution within this population was as
follows: 16.2\% were aged 20 to 29 years, 57.8\% were aged 30 to 64
years, and 25.9\% were aged 65 to 80 years. From this population, 500
individuals were randomly selected and contacted, out of which 425
participated in the study. A total of 242 individuals with confirmed
infection had been hospitalized since the beginning of the outbreak,
and 59 people with confirmed infection had passed away.

The study revealed a population-wide measured prevalence $\hat f_{1\land 2}^*$ of 
1.9\% (0.8 -- 3.7\%). Seroprevalence was assessed using two different
assays: (i) Abbott SARS-CoV-2 IgG and (ii) Euroimmun Anti-SARS-CoV-2
ELISA (IgG). The former has a sensitivity and specificity of 83.1\%
(75.4 -- 100\%) and 100\%,
respectively~\cite{castro2021performance}. The sensitivity and
specificity of the latter are 91.1\% (80.7 -- 96.1\%) and 100\% (96.5
-- 100\%), respectively~\cite{fda2024}.

Every individual who tested positive in Abbott's assay underwent
confirmation using Euroimmun's Anti-SARS-CoV-2 ELISA (IgG). This
process represents an $\mathrm{AND}$ aggregation.
\section*{Acknowledgments}
The authors thank Stefan Felder and Nana Owusu-Boaitey for helpful comments. We acknowledge financial support 
from hessian.AI (LB), the ARO through grant W911NF-23-1-0129 (MRD and LB), and 
the NSF through grant OAC-2320846 (MRD).
\nolinenumbers

% Either type in your references using
% \begin{thebibliography}{}
% \bibitem{}
% Text
% \end{thebibliography}
%
% or
%
% Compile your BiBTeX database using our plos2015.bst
% style file and paste the contents of your .bbl file
% here. See http://journals.plos.org/plosone/s/latex for 
% step-by-step instructions.
% 

\bibliography{refs}

\end{document}